\documentclass[aps,twocolumn,showpacs,preprintnumbers,prb,superscriptaddress]{revtex4-2}

\usepackage[english]{babel}

\usepackage{graphics}
\usepackage{color}
\usepackage{amssymb}
\usepackage{amsmath}
\usepackage{overpic}
\usepackage{epstopdf}
\usepackage{threeparttable}
\usepackage{lipsum}
\usepackage{bm}
\usepackage{braket}
\usepackage{graphicx}
\usepackage{subfigure}
\usepackage{multirow}
\usepackage{array}
\usepackage{newtxtext}
\usepackage[dvipsnames]{xcolor}

\usepackage{hyperref}
\hypersetup{plainpages=false,colorlinks=true,linkcolor=blue, citecolor=blue, urlcolor=blue}

\usepackage{longtable}
\usepackage{booktabs}
\usepackage{blindtext}

\begin{document}

\title{Lattice dynamics and its effects on magnetocrystalline anisotropy energy of pristine and hole-doped YCo$_5$ from first principles}

\author{Guangzong Xing}
\email{XING.Guangzong@nims.go.jp}
\affiliation{Research center for Magnetic and Spintronic Materials, National Institute for Materials Science, Tsukuba, Ibaraki, 305-0047, Japan}
\affiliation{Elements Strategy Initiative Center for Magnetic Materials, National Institute for Materials Science, Tsukuba, Ibaraki, 305-0047, Japan}

\author{Yoshio Miura}
\affiliation{Research center for Magnetic and Spintronic Materials, National Institute for Materials Science, Tsukuba, Ibaraki, 305-0047, Japan}

\author{Terumasa Tadano}
\email{TADANO.Terumasa@nims.go.jp}
\affiliation{Research center for Magnetic and Spintronic Materials, National Institute for Materials Science, Tsukuba, Ibaraki, 305-0047, Japan}
\affiliation{Elements Strategy Initiative Center for Magnetic Materials, National Institute for Materials Science, Tsukuba, Ibaraki, 305-0047, Japan}
\date{\today}

\begin{abstract}
We study the lattice dynamics effects on the phase stability and magnetocrystalline anisotropy (MCA) energy of CaCu$_5$-type YCo$_5$ at finite temperatures using first-principles calculations based on density functional theory (DFT). Harmonic lattice dynamics (HLD) calculations indicate that YCo$_5$ with 56 full valance electrons is dynamically unstable and this instability can be cured by reducing the number of electrons ($N_e$). Crystal orbital Hamilton population analysis reveals that the observed phonon instability originates from the large population of antibonding states near the Fermi level, which is dominated by the Co atoms in the honeycomb layer. The antibonding state depopulates with decreasing $N_e$, resulting in stable phonons for hole-doped YCo$_5$ with $N_e$ $\leq$ 55. We then evaluate the temperature-dependent MCA energy using both HLD and \textit{ab initio} molecular dynamics (AIMD) methods. For the pristine YCo$_5$, we observe a very weak temperature decay of the MCA energy, indicating little effect of lattice dynamics. Also, the MCA energies evaluated with AIMD at all target temperatures are larger than that of the static hexagonal lattice at 0 K, which is mainly attributed to the structural distortion driven by soft phonon modes. In the hole-doped YCo$_5$, where the distortion is suppressed, a considerable temperature decay in MCA energy is obtained both in HLD and AIMD methods, showing that lattice dynamics effects on MCA energy are non-negligible.

\end{abstract}

\maketitle

\section{Introduction}

The intermetallic compound CaCu$_5$-type RECo$_5$ (RE= rare earth) has been extensively studied in the past, both experimentally and theoretically, due to its superior magnetic properties such as high Curie temperature, saturation magnetization, and strong coercivity~\cite{RECo5,RECo5_1,RECo5_2,RECo5_3,RECo5_4}. One of the intrinsic magnetic properties associated with high coercivity is magnetocrystalline anisotropy (MCA) energy, which depends only on the crystal structures and chemical compositions. MCA energy is defined as the ground state energy difference between different directions of the magnetic field with respect to the crystal axes. In the previous studies, it was found both experimentally and theoretically that RECo$_5$ (RE = Y, La, Ce, Sm) exhibits a uniaxial MCA energy that results in an energetically favored alignment of the magnetic moments along the crystallographic $c$-axis~\cite{YCo5_mae_exp,YCo5_mae_GGAU,CeCo5,CeCo5_1,LaCo5,SmCo5,SmCo5_mae_exp}. In RECo$_5$, the large MCA mainly arises from two aspects: the spin-orbit coupling of the itinerant $3d$ electrons at the Co sites, and the spin-orbit interaction of the localized $4f$ electrons at the RE site. Compared with $3d$ electrons, the MCA originating from $4f$ electrons shows a strong temperature dependence. The two contributions become comparable at finite temperatures. For example, Alameda \textit{et al.} reported the MCA constant of YCo$_5$ was 7.4 MJ/m$^3$ at 4.2~K and slightly decreased to 5.8 MJ/m$^3$ at room temperature~\cite{YCo5_mae_exp}. The corresponding values for the well-known SmCo$_5$ are 30 MJ/m$^3$ and 17 MJ/m$^3$~\cite{SmCo5_mae_exp}, respectively.

A theoretical study of the temperature effect on MCA energy is crucial since permanent magnets are usually used in a high-temperature environment. An important physical factor affecting MAC at finite temperature is the thermal fluctuation of the spin moment. This effect has approximately been included in density functional theory (DFT) calculations with disordered local moment (DLM) approach~\cite{DLM_appro}. For example, Matsumoto \textit{et al.}~\cite{YCo5_mae_DLM1} investigated the temperature dependence of MCA energy and magnetization in hole-doped YCo$_5$ using the DLM approach. A DLM-based first-principles magnetization versus field (FPMVB) approach is introduced by Patrick \textit{et al.} to study the temperature-dependent MCA energy of YCo$_5$ and GdCo$_5$~\cite{YCo5_mae_DLM}. They found excellent agreement of temperature-dependent MCA energy curves in GdCo$_5$ between the FPMVB approach and experiments. 

Lattice dynamics is another important factor that can affect the MCA energy at finite temperature~\cite{MAE_ave_MnBi,MnBi_K_vib}. At elevated temperatures, thermal excitation of phonons is expected to change the electronic structures, which then influences the MCA energy. For example, Urru \textit{et al.}~\cite{MnBi_K_vib} has recently reported that the spin-reorientation transition 
of MnBi could be explained by considering the vibrational free energy contribution calculated within the harmonic approximation (HA). Thus, it is intriguing to study lattice dynamics effects on the temperature-dependent MCA energy of permanent magnetic materials, particularly the CaCu$_5$-type RECo$_5$ which displays the same hexagonal lattice as MnBi.

In this work, we theoretically investigate the lattice dynamics and its effect on the MCA energy of YCo$_{5-x}$ at finite temperature to study the MCA energy originating from the itinerant states. By performing phonon calculations of YCo$_5$ within the harmonic approximation, we show that the experimentally-reported CaCu$_5$ structure is dynamically unstable. This phonon instability originates from the presence of large antibonding states near the Fermi level, which can be removed by reducing the number of electrons ($N_e$). We find that at least one electron needs to be removed from the system to stabilize phonons of YCo$_{5-x}$. For the hole-doped YCo$_{5-x}$, we then evaluate the lattice dynamics effect on the MCA energy from the difference of the vibrational free energies computed with different spin orientations. In the high-temperature region, the phonon contribution to the MCA energy becomes significant and comparable to the MCA energy at 0 K obtained from DFT calculation. To evaluate the lattice dynamics effect in the undoped YCo$_5$, we also perform \textit{ab initio} molecular dynamics (AIMD) simulations and find that the MCA energy hardly changes with increasing temperature. We attribute this weak temperature dependence mainly to structural distortion.

The structure of this paper is organized as follows. In the next section, we describe our theoretical methods in detail. The MCA energy both at 0~K and finite temperatures, and computational details are described in Secs.~\ref{subsec:MAE_0K}, \ref{subsec:MAE_finite}, and \ref{subsec:computational_detail}, respectively.  In Sec.~\ref{sec:results}, we show our main results. The dynamical instability of the CaCu$_5$-type YCo$_5$ with 56 full valance electrons and the microscopic origin of stabilization by reducing $N_e$, or hole-doping, are discussed in Sec.~\ref{subsec:dynamical_stability}. In Sec.~\ref{subsec:MAE_constant}, we evaluate the MCA energy both at 0~K and finite temperatures using different approaches and discuss the role of lattice dynamics in the temperature dependence of the MCA energy. Finally, we summarize this study in Sec.~\ref{sec:summary}. 

\section{Methods}

\label{sec:method}

\subsection{MCA energy at 0~K from DFT calculation}
\label{subsec:MAE_0K}
We perform a noncollinear spin-orbit interaction calculation using the well-known force theorem~\cite{force_theorem} to obtain the MCA energy, $K_\mathrm{u}^{\mathrm{DFT}}$ at 0~K, which is defined as
\begin{equation}
    K_{\mathrm{u}}^{\mathrm{DFT}} = E_{\perp}^{\mathrm{DFT}}-E_{\parallel}^{\mathrm{DFT}},\label{eq:K_DFT}
\end{equation}
where $E_{\perp}^{\mathrm{DFT}}$ and $E_{\parallel}^{\mathrm{DFT}}$ are the sum of the occupied Kohn--Sham eigenenergies with the magnetic moment ($\bm{m}$) being aligned along the hard ([100] direction, $\bm{m}\perp \bm{c}$) and easy ([001] direction, $\bm{m}\parallel\bm{c}$) axes, respectively. The positive $K_{\mathrm{u}}^{\mathrm{DFT}}$ indicates that the energetically favorable $\bm{m}$ is parallel to the crystallographic $\bm{c}$-axis.

In order to understand the atomic site-dependent MCA energy, we carry out the second-order perturbation calculation. In the tight-binding regime, the Hamiltonian for spin-obit coupling is given by the sum of the contributions from each atomic site, $H_{\mathrm{SO}}=\sum_i\xi_i\bm{L}_i\cdot\bm{S}_i$, where $\bm{L}_i (\bm{S}_i)$ is the single-electron angular (spin) momentum operator, and $\xi_i$ is the spin-orbit coupling constant of atom $i$. Compared with 3$d$ bandwidth, $\xi_i$ ($\xi_\mathrm{Co}$ = 69.4~meV) is relatively small, which can be treated as a perturbation term. The second-order  perturbation energy is expressed as
\begin{equation}
 E^{(2)}=-\sum_{\bm{k}}\sum_{n^{\prime}\sigma^{\prime}}^{\mathrm{unocc}}\sum_{n\sigma}^{\mathrm{occ}}\frac{|\langle\bm{k}n^{\prime}\sigma^{\prime}|H_{\mathrm{SO}}|\bm{k}n\sigma\rangle|^2}{\epsilon_{\bm{k}n^{\prime}\sigma^{\prime}}^{(0)}-\epsilon_{\bm{k}n\sigma}^{(0)}}, \label{eq:Sec_order}  
\end{equation}
where $|\bm{k}n\sigma\rangle$ is the unperturbed state with energy $\epsilon_{\bm{k}n\sigma}^{(0)}$. $\bm{k}$, $n$, and $\sigma$ represent the wave vector, band index, and spin, respectively. The index "occ" and "unocc" means the sum over the occupied and unoccupied states. 
$|\bm{k}n\sigma\rangle$ can be expanded as a sum of atomic  orbitals, $|\bm{k}n\sigma\rangle=\sum_{i\mu}c_{i\mu\sigma}^{\bm{k}n}|i\mu\sigma\rangle$, where the atomic orbitals labeled as $\mu$ and the coefficient $c_{i\mu\sigma}^{\bm{k}n}$ can be obtained by DFT calculations. Therefore, the site-dependent second-order contribution to the total energy with different spin processes and atomic orbitals can be calculated using first-principles calculations. 

The MCA energy at 0~K, $K_{\mathrm{u}}^{\mathrm{PT}}$, within the second-order perturbation is defined as $K_{\mathrm{u}}^{\mathrm{PT}} = E_{\perp}^{(2)}-E_{\parallel}^{(2)}$, where $E_{\perp}^{(2)}$ ($E_{\parallel}^{(2)}$) are total energies calculated by Eq.~(\ref{eq:Sec_order}) with the magnetization along hard (easy) axis of CuCa$_5$-type YCo$_5$. Then the decomposed part of $K_{\mathrm{u}}^{\mathrm{PT}}$ at different atomic site with spin-transition process is written as 
\begin{align}
    K_{\mathrm{u}}^{\mathrm{PT}} & = \sum_i{}K_{\sigma\Rightarrow\sigma^{\prime}}^i \notag \\
    & = \sum_i{}K_{\uparrow\Rightarrow\uparrow}^{i}+K_{\downarrow\Rightarrow\downarrow}^{i}+K_{\uparrow\Rightarrow\downarrow}^{i}+K_{\downarrow\Rightarrow\uparrow}^{i}. \label{eq:K_sec}
\end{align}
The first two terms in Eq.~(\ref{eq:K_sec}) are called spin-conserving terms and the rest are spin-flip terms, which originate from the spin scattering process between the occupied and unoccupied state near the Fermi level. The detailed formulation of $K_{\sigma\Rightarrow\sigma^{\prime}}^i$ is given in Ref.~\cite{MAE_sec}.

\subsection{Lattice dynamics contribution to MCA energy at finite temperatures}
\label{subsec:MAE_finite}

The MCA energy at finite temperatures, $K_{\mathrm{u}}$, can be evaluated by the difference of Helmholtz free energies computed with two different spin orientations, which is defined as
\begin{equation}
    K_{\mathrm{u}}(V,T) = F_{\perp}(V,T)-F_{\parallel}(V,T), \label{eq:free}
\end{equation}
where $V$ is the volume of YCo$_5$. According to adiabatic approximation, $K_{\mathrm{u}}$ is considered as the sum of the following terms

\begin{align}
    K_{\mathrm{u}}& \approx {}K_{\mathrm{u}}^{\mathrm{DFT}}(V_0)+K_{\mathrm{u}}^{\mathrm{el}}(V_0,T)+K_{\mathrm{u}}^{\mathrm{phon}}(V_0,T) \notag\\
    &\hspace{30mm}+K_{\mathrm{u}}^{\mathrm{mag}}(V_0,T)+K_{\mathrm{u}}^{\mathrm{TE}}(T), \label{eq:K_u}
   \end{align}
where $K_{\mathrm{u}}^{\mathrm{DFT}}$($V_0$) is the MCA energy obtained from the DFT calculation with the optimized volume $V_0$, $K_{\mathrm{u}}^{\mathrm{el}}$($V_0,T$) and $K_{\mathrm{u}}^{\mathrm{phon}}$($V_0,T$) represent the electronic and vibrational contribution to $K_{\mathrm{u}}$, respectively. $K_{\mathrm{u}}^{\mathrm{mag}}(V_0,T)$ is the magnon contribution, which is not considered in the current study. $K_{\mathrm{u}}^{\mathrm{el}}$ for the present system is very small and thus can be ignored. We roughly evaluated the thermal expansion term, $K_{\mathrm{u}}^{\mathrm{TE}}(T) \approx K_{\mathrm{u}}^{\mathrm{DFT}}(V(T))-K_{\mathrm{u}}^{\mathrm{DFT}}(V_0)$, using the measured temperature-dependent lattice constants of YCo$_5$ from Ref.~\cite{YCo5_thermal}. Since the change in the lattice constant $c$ is negligible, we only change the in-plane lattice constant $a$ as $a(T) = kT+a_0$, where $a_0$ is the lattice constant optimized by DFT at $T = 0$ K, and $k\sim 7\times10^{-5}$ \AA\, K$^{-1}$ is the proportionality constant estimated from the experimental temperature dependence of the $a$ value~\cite{YCo5_thermal}.

The vibrational MCA energy $K_{\mathrm{u}}^{\mathrm{phon}}$ is computed from the difference between the vibrational free energies at different magnetic orientations. In the HA, it becomes~\cite{MnBi_K_vib}

\begin{equation}
    K_{\mathrm{u}}^{\mathrm{phon}} = -\frac{1}{\beta}\ln{\frac{Z_{0,\perp}}{Z_{0,\parallel}}}
    \label{eq:K_vib}
\end{equation}
where $\beta=(k_{\mathrm{B}}T)^{-1}$ is the inverse temperature with $k_{\mathrm{B}}$ being the Boltzmann constant, and $Z_{0,\bm{m}}$ is the partition function of the harmonic oscillator at the magnetic state $\bm{m}$ defined as
\begin{equation}
    Z_{0,\bm{m}} = \prod_{\bm{q}\nu}\frac{1}{2\sinh{\left(\frac{\beta\hbar\omega_{\bm{q}\nu,\bm{m}}}{2}\right)}},
    \label{eq:Z0}
\end{equation}
with $\hbar$ being the reduced Planck constant. $\omega_{\bm{q}\nu,\bm{m}}$ is the harmonic phonon frequency of the $\nu$th branch at crystal momentum $\bm{q}$ in the magnetic state $\bm{m}$, which can be obtained by diagonalizing the dynamical matrix constructed from the second-order interatomic force constants (IFCs) calculated in the same magnetic state $\bm{m}$. If the difference of $\Delta \omega_{\bm{q}\nu} = \omega_{\bm{q}\nu,\perp} - \omega_{\bm{q}\nu,\parallel}$ is small, Eq.~(\ref{eq:K_vib}) can be approximated by a linear function of $\Delta\omega_{\bm{q}\nu}$ as 
\begin{equation}
    K_{\mathrm{u}}^{\mathrm{phon}} \approx \frac{\hbar}{2}\sum_{\bm{q}\nu} [1 + 2n(\omega_{\bm{q}\nu,\parallel})] \Delta \omega_{\bm{q}\nu},
    \label{eq:K_phon}
\end{equation}
where $n(\omega) = (e^{\beta\hbar\omega}-1)^{-1}$ is the Bose--Einstein distribution function. Since $\Delta \omega_{\bm{q}\nu}$ is very small in many cases, Eq.~(\ref{eq:K_phon}) is a reasonable approximation to Eq.~(\ref{eq:K_vib}) and will be used for computing the modal contribution to $K_{\mathrm{u}}^{\mathrm{phon}}$ in Sec.~\ref{subsub:MAE_finite}. In the classical limit ($\hbar\rightarrow 0$), Eq.~(\ref{eq:K_phon}) reduces to $K_{\mathrm{u}}^{\mathrm{phon}}\approx\beta^{-1}\sum_{\bm{q}\nu}(\Delta \omega_{\bm{q}\nu}/\omega_{\bm{q}\nu,\parallel})$.

It is important to recall that the above harmonic lattice dynamics (HLD) method can be employed only when the structure is dynamically stable, namely, $\omega_{\bm{q}\nu}\geq 0$ is satisfied for all phonon modes in the Brillouin zone. If the structure is dynamically unstable, which is the case for pristine YCo$_5$ as will be shown later, Eq.~(\ref{eq:Z0}) becomes ill-defined and the HLD method breaks down. Hence, a beyond (quasi-)harmonic approach is required for evaluating the vibrational contribution to the MCA energy of YCo$_{5}$. The limitation of the HLD method can be overcome by using anharmonic lattice dynamics methods, such as the self-consistent phonon theory~\cite{Errea_PRB2014,ALAMODE_1,Oba_PRM2019} or temperature-dependent effective potential method~\cite{Hellman_PRB2013}, or \textit{ab initio} molecular dynamics (AIMD) method. 
In this study, we employ AIMD and evaluate the finite-temperature MCA energy as
\begin{equation}
    \Braket{K_{\mathrm{u}}}_{\mathrm{MD}}(T) = \frac{1}{N}\sum_{s=1}^N K_{\mathrm{u}}^{\mathrm{DFT}} (\{\bm{R}_i\}_{s}),\label{eq:K_md}
\end{equation}
where $K_{u}^{\mathrm{DFT}}(\{\bm{R}_i\}_{s})$ is the MCA energy computed at the $s$th structure snapshot $\{\bm{R}_i\}_{s}$ sampled from the AIMD trajectory at the target temperature $T$, and $N$ is the number of structures sampled at each $T$. This approach can be used even for the systems where an unstable phonon mode ($\omega_{\bm{q}\nu}^2<0)$ exits. Besides, contributions from the anharmonic terms of the potential energy surface and the effect of structural distortion are included automatically.

\subsection{Computational details}
\label{subsec:computational_detail}

First-principle calculations in this study were performed by using the projector augmented wave (PAW) method~\cite{paw}, within the Perdew--Burke--Ernzerhof (PBE) generalized gradient approximation (GGA)~\cite{pbe-gga}, as implemented in the Vienna \textit{ab initio} simulation package (VASP)~\cite{vasp}. Lattice constants and atomic positions of CaCu$_5$-type YCo$_5$ were carefully optimized with a kinetic-energy cutoff of 400 eV for the plane-wave expansion, and the $k$-point mesh was generated automatically in such a way that the mesh density in the reciprocal space became larger than 450 \AA$^{-3}$. For the structural optimization and phonon calculations, we used the Methfessel--Paxton smearing method~\cite{Methfessel_Paxton1989} with the width of 0.2 eV. On the other hand, the Methfessel--Paxton smearing method with a smaller width of 0.05 eV and tetrahedron method with the Bl\"{o}chl correction~\cite{Blochl_PRB1994} was used for calculating the MCA energy. Note that $k$ mesh density of $\sim$6000 \AA$^{-3}$ was used to calculate $K_{\mathrm{u}}^{\mathrm{DFT}}$ ($K_{\mathrm{u}}^{\mathrm{PT}}$) because the anisotropy energy is small, necessitating a denser mesh. The initial local moments of 3 $\mu_B$ and $-0.3$ $\mu_B$ were set for Co and Y atoms, respectively, for the collinear spin-polarized calculations, including phonon calculations. 

The harmonic phonon calculations were conducted by using the finite-displacement method, as implemented in \textsc{alamode}~\cite{ALAMODE,ALAMODE_1}. A $2\times2\times2$ supercell containing 48 atoms was adopted for the phonon calculation. To obtain  $K_{\mathrm{u}}^{\mathrm{phon}}$ using Eq.~(\ref{eq:K_vib}),
we incorporated noncollinear spin-orbit interaction into the DFT calculation and estimated the second-order IFCs for two different spin orientations: $\bm{m}\perp\bm{c}$ and $\bm{m}\parallel\bm{c}$. We have confirmed that our implementation yields consistent results with the density-functional perturbation theory implementation~\cite{MnBi_K_vib} for MnBi. In addition, AIMD simulations were carried out to evaluate $\Braket{K_{\mathrm{u}}}_{\mathrm{MD}}(T)$ using Eq.~(\ref{eq:K_md}). 
At each temperature, we performed a collinear AIMD run for 5000 steps with a time-step of 2~fs and extracted 100 structural snapshots uniformly from the last 2500 steps. For each sampled snapshot, we then calculated the difference of the total energies between the two spin orientations. Finally, we estimated the average over the 100 snapshots using Eq.~(\ref{eq:K_md}). 
We have confirmed that 100 structures were sufficient to obtain converged values of $\Braket{K_{\mathrm{u}}}_{\mathrm{MD}}$.

\section{Results and Discussion}
\label{sec:results}

\begin{figure}[bthp]
 \centering
 \includegraphics[width=0.46\textwidth]{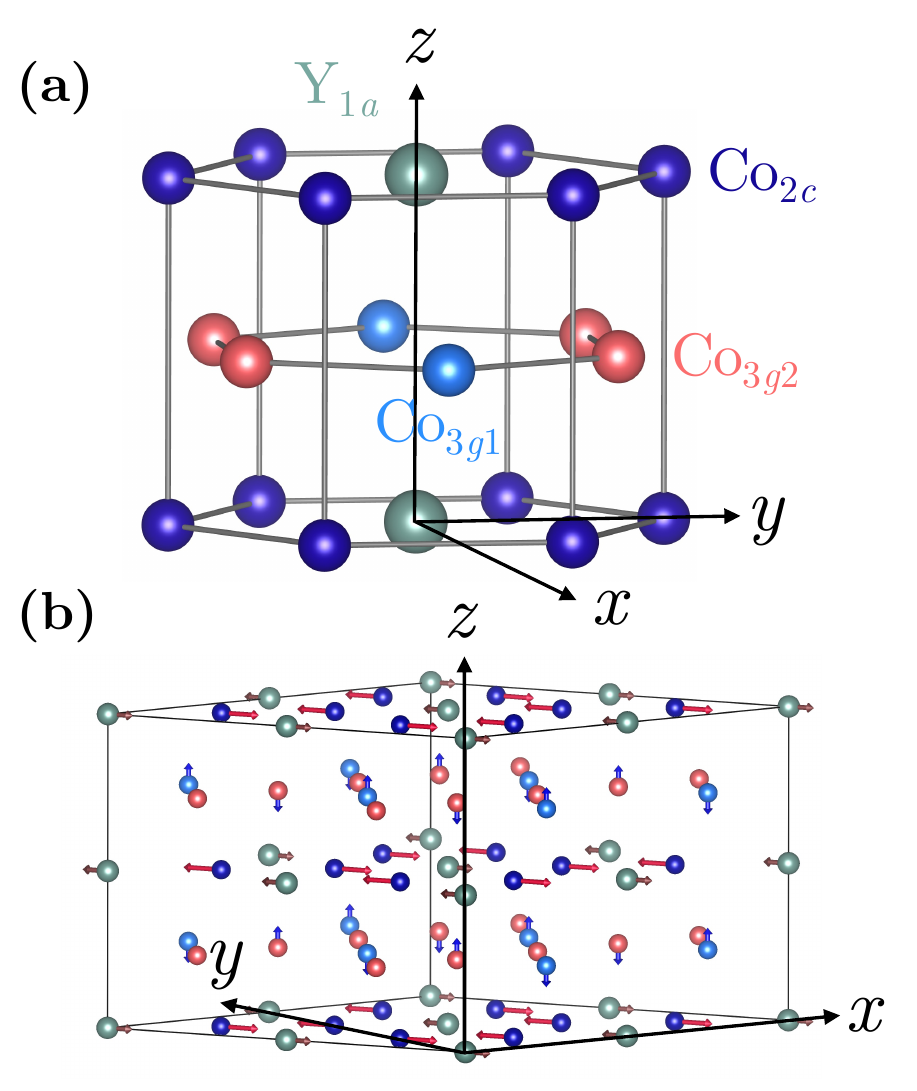}
 \caption{(a) Crystal structure of CaCu$_5$-type YCo$_5$. The large spheres represent Y atoms, and the smaller ones are Co atoms. Wyckoff positions for different atoms are marked with the corresponding atom colors. (b) Crystal structure of YCo$_5$ with a 2$\times$2$\times$2 supercell. The arrows indicate the displacement pattern of the the lowest-energy soft phonon mode at L point [$\bm{q}=$($\frac{1}{2}$, 0, $\frac{1}{2}$)]. The length of the arrows represents the relative magnitude of displacements. Cartesian axes are labeled as $x$, $y$, and $z$, respectively.}
 \label{fig:disp_pattern}
\end{figure}

\begin{figure*}
 \centering
 \includegraphics[width=0.8\textwidth]{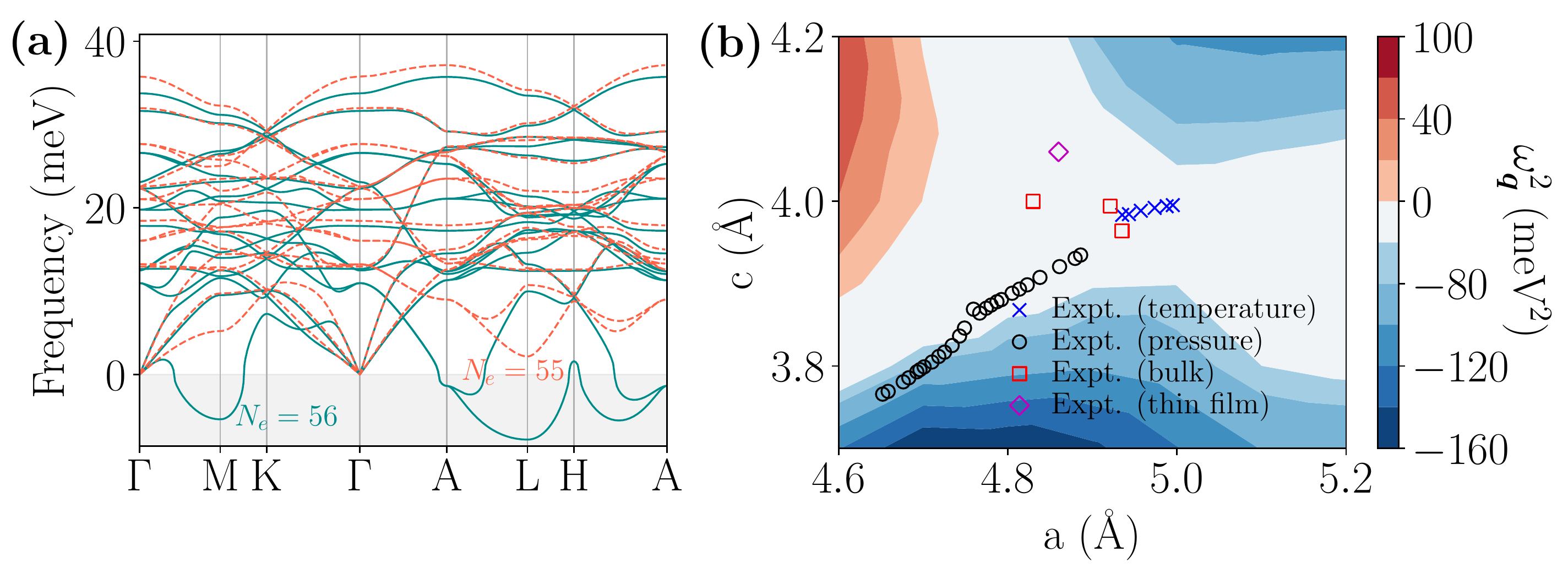}
     \caption{Phonon frequencies of CaCu$_5$-type YCo$_5$ computed within harmonic approximation. (a) Phonon dispersion curves for the pristine YCo$_5$ ($N_e=56$, solid line) and the hole-doped system ($N_e=55$, dashed line). (b) Contour plot of the smallest $\omega_{\bm{q}}^2$ value for YCo$_5$ ($N_e$ = 56) computed with various lattice parameters. The experimental lattice parameters are shown with open symbols. Circles and crosses are temperature- and hydrostatic pressure-dependence of lattice parameters from Ref.~\cite{YCo5_thermal} and Ref.~\cite{YCo5_latt_exp4}, respectively. Squares and diamonds are lattice parameters of bulk~\cite{YCo5_latt_exp,YCo5_latt_exp1,YCo5_latt_exp2} and thin-film phases~\cite{YCo5_latt_exp3}, respectively.}
 \label{fig:phonon-main}
\end{figure*}

CaCu$_5$-type YCo$_5$ displays a layered hexagonal structure (space group: $P6/mmm$) shown in Fig.~\ref{fig:disp_pattern}(a). The cobalt atoms are located either at the $2c$ or $3g$ Wyckoff sites, which respectively belong to the honeycomb and kagome layers. We label these inequivalent cobalt atoms as Co$_{2c}$ or Co$_{3g}$.
The symmetry of YCo$_5$ will be lowered with spin-orbit interaction when the magnetization is along the hard axis ([100] direction). The 3-fold Co$_{3g}$ will be separated into two inequivalent cites, which have the corresponding multiplicities of 1 (denoted as $3g1$ hereafter) and 2 (denoted as $3g2$ hereafter), respectively. The optimized lattice constants are $a=4.907$ \AA{} and $c=3.942$ \AA{}, which agree reasonably well with the experimental values at 5 K ($a=4.936$ \AA, $c=3.984$ \AA, Ref.~\cite{YCo5_thermal}) and at room temperature ($a=4.950$ \AA, $c=3.986$ \AA, Ref.~\cite{YCo5_thermal}; $a=4.921$ \AA, $c=3.994$ \AA, Ref.~\cite{YCo5_latt_exp}). 

\subsection{Dynamical stability}
\label{subsec:dynamical_stability}

\begin{figure*}
 \centering
 \includegraphics[width=0.80\textwidth]{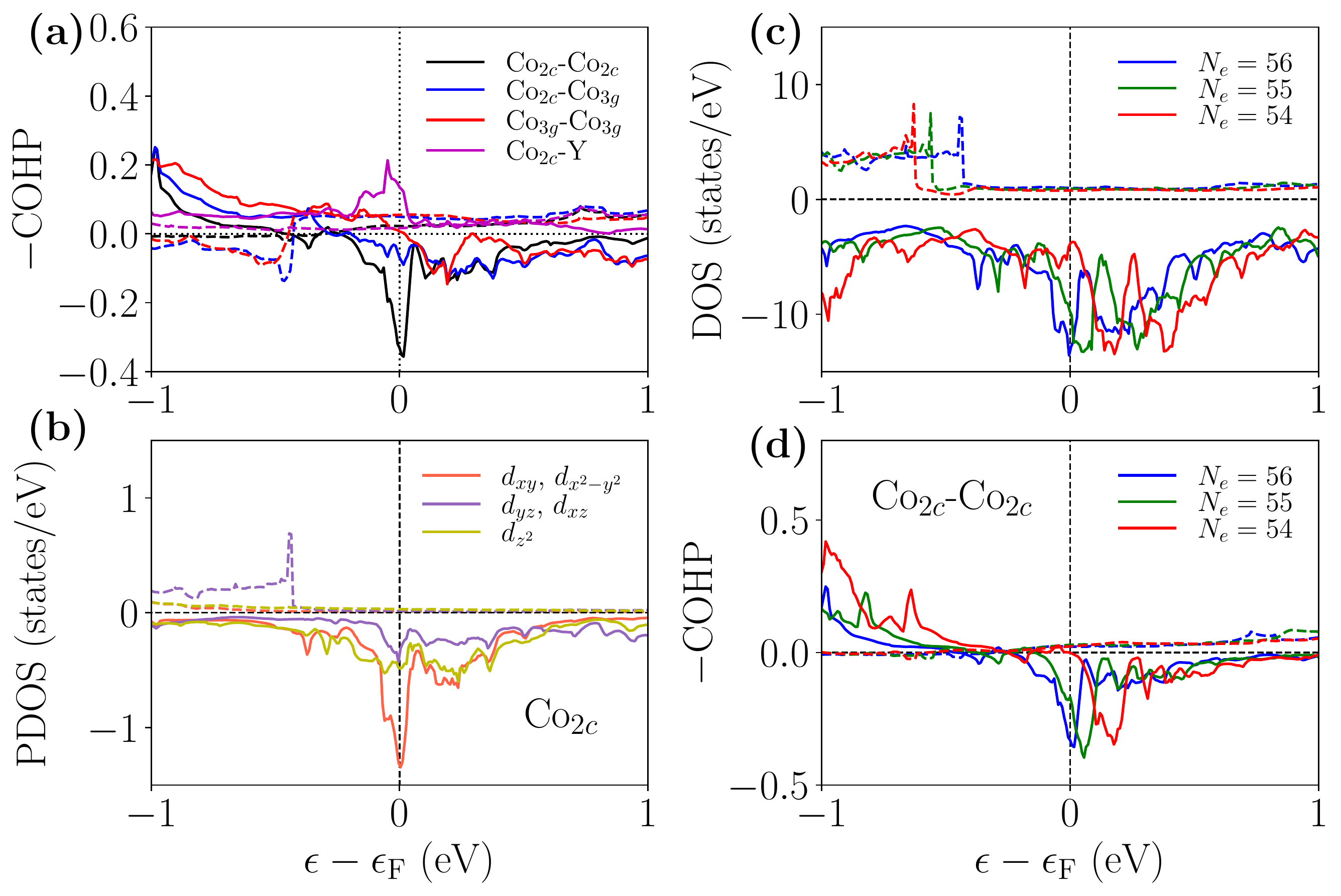}
 \caption{(a) COHP calculation of different atomic bonds for YCo$_5$ with $N_e=56$. (b) Projected density of states (PDOS) for 3$d$ states at Co$_{2c}$ site. (c) The total density of states (DOS) and (d) COHP calculation of Co$_{2c}$-Co$_{2c}$ bonds for YCo$_5$ with $N_e$ = 56, 55, and 54 respectively. The solid and dash lines in all panels represent minority- and majority-spin, respectively. }
 \label{fig:DOS-COHP}
\end{figure*}

The harmonic phonon dispersion curves of the CaCu$_5$-type YCo$_5$ are shown in Fig.~\ref{fig:phonon-main}(a) by solid lines.
While the previous experimental studies showed that the crystal structure of YCo$_5$ is the CaCu$_5$ type at and above room temperature, the harmonic phonon of the pristine YCo$_5$ turns out to be unstable. 
The largest phonon instability occurs at the L points of $\bm{q}=$($\frac{1}{2}$, 0, $\frac{1}{2}$), (0, $\frac{1}{2}$, $\frac{1}{2}$), and ($\frac{1}{2}$, $\frac{1}{2}$, $\frac{1}{2}$). The second-largest instability is observed at the M points of $\bm{q}=$($\frac{1}{2}$, 0, 0), (0, $\frac{1}{2}$, 0), and ($\frac{1}{2}$, $\frac{1}{2}$, 0).
Since the calculated phonon frequency is rather sensitive to the lattice parameters, we also computed the phonon dispersion curves with various $a$ and $c$ values. The results are shown in Fig.~S1 of the supplementary material (SM)~\cite{supplement}. As shown in Fig.~\ref{fig:phonon-main}(b), the smallest $\omega_{\bm{q}}^2$ value, which appears either at the L point or M point, is negative in the wide region of the $ac$ plane, including the area around the experimental $(a,c)$ values reported in  Refs.~\cite{YCo5_thermal,YCo5_latt_exp,YCo5_latt_exp1,YCo5_latt_exp2,YCo5_latt_exp3,YCo5_latt_exp4}.  While the phonons can be 
dynamically stable in the upper left region ($a \lesssim 4.7$~\AA{}, $c \gtrsim 3.9$~\AA) of the figure, the experimental and theoretical lattice parameters are located away from that region. Thus, the observed dynamical instability of YCo$_5$ cannot exclusively be attributed to the slight errors in the GGA-PBE lattice parameters.

The soft modes at the L point and M point involve the in-plane displacements of Co$_{\mathrm{2c}}$ atoms accompanied by relatively small displacements of other atoms, as shown in Fig.~\ref{fig:disp_pattern}(b). Once the atoms are slightly displaced along the polarization vector of the soft mode, the six-fold rotational symmetry in the honeycomb layer breaks, which decreases the total energy of the system. To understand the origin of the phonon instability, we investigated the bonding nature of the nearest-neighbor atomic pairs by computing the Crystal orbital Hamilton Populations (COHPs)~\cite{COHP} using the \textsc{lobster} code~\cite{COHP1}. The $-$COHP values computed with collinear magnetism are shown in Fig.~\ref{fig:DOS-COHP}(a), where the dash and solid lines represent the results for the majority and minority spins, respectively. The most notable feature in Fig.~\ref{fig:DOS-COHP}(a) is the large negative peak of $-$COHP for the nearest Co$_{2c}$-Co$_{2c}$ bond at the Fermi level, representing its strong antibonding nature. The projected DOS in Fig.~\ref{fig:DOS-COHP}(b) shows that the antibonding state is formed by the $d_{xy}$ and $d_{x^{2}-y^{2}}$ orbitals, while the contributions from the other orbitals both at Co$_{2c}$ and Co$_{3g}$ (see Fig.~S2 in the SM~\cite{supplement}) sites are negligible around the Fermi level. The large antibonding contribution is energetically unfavorable. Hence, the system reacts in such a way that the population of the antibonding state decreases at the Fermi level, which can be achieved by distorting the structure and thereby lifting the orbital degeneracy.

The large antibonding population at the Fermi level may be reduced by hole doping, which is expected to improve the stability of the CaCu$_5$ structure. To see the effect of hole doping, we gradually reduced the number of valence electrons, $N_e$, from the original value of the pristine YCo$_5$ ($N_e = 56$) and computed the $-$COHP values for the relevant Co$_{2c}$-Co$_{2c}$ bond. As shown in Fig.~\ref{fig:DOS-COHP}(c), the Fermi level lowers with reducing $N_e$, and the peak position of the DOS in the minority spin state shifts to the higher energy. In addition, the population of the Co$_{2c}$-Co$_{2c}$ antibonding state at the Fermi level decreases as decreasing $N_e$. Consequently, we observed that the phonons of the CaCu$_5$ structure became dynamically stable when $N_e \lesssim 55$ (see Fig.~\ref{fig:phonon-main}(a) for the $N_e=55$ case) and the soft-mode frequency at the L point increased gradually with decreasing $N_e$, as tabulated in Table \ref{table:MAE}.

\begin{table}[bth]
\centering
  \caption{Calculated $N_e$-dependence of the MCA energies ($K_{\mathrm{u}}^{\mathrm{DFT}}$, $K_{\mathrm{u}}^{\mathrm{PT}}$, $K_{\mathrm{u}}^{\mathrm{VCA}}$) using different approaches (see the main text), anisotropy in the orbital moment ($\Delta{}m_{\mathrm{o}}$), and the squared frequency of the soft mode at the L point ($\omega_{\mathrm{L}}^2$).}
  \label{table:MAE}
  \small
  \begin{ruledtabular}
  \begin{tabular}{lrrrrr}
    \multirow{3}{*}{$N_e$} &\multicolumn{3}{c}{MCA constant (meV/f.u.)} &\multirow{3}{*}{$\Delta{}m_{\mathrm{o}}$ ($\mu_{\mathrm{B}}$/f.u.)}    &\multirow{3}{*}{$\omega_{\mathrm{L}}^2$ (meV$^2$)}\\
    \cline{2-4}
     &\multirow{2}{*}{$K_{\mathrm{u}}^{\mathrm{DFT}}$} &\multirow{2}{*}{$K_{\mathrm{u}}^{\mathrm{PT}}$}  &\multirow{2}{*}{$K_{\mathrm{u}}^{\mathrm{VCA}}$}  &     &\\\\
    \toprule[0.15mm]
     $56$ & $0.383$ & $0.177$ & $0.383$ &$0.074$ &$-61.1$\\
     $55$ & $2.947$ & $2.451$ & $2.757$ &$0.206$&  $4.7$\\
     $54$ & $1.797$ & $1.442$ &$-0.611$ &$0.153$&  $79.3$\\
     $53$ & $0.356$ & $0.426$ &$-1.407$ &$0.074$& $130.1$\\
     $52$ &$-0.643$ &$-0.384$ &$-1.553$ &$0.028$& $156.3$\\
     $51$ &$-1.069$ &$-0.833$ &$-1.331$ &$0.001$& $156.1$\\
  \end{tabular}
  \end{ruledtabular}
\end{table}

\begin{figure}[!tb]
 \centering
 \includegraphics[width=0.46\textwidth]{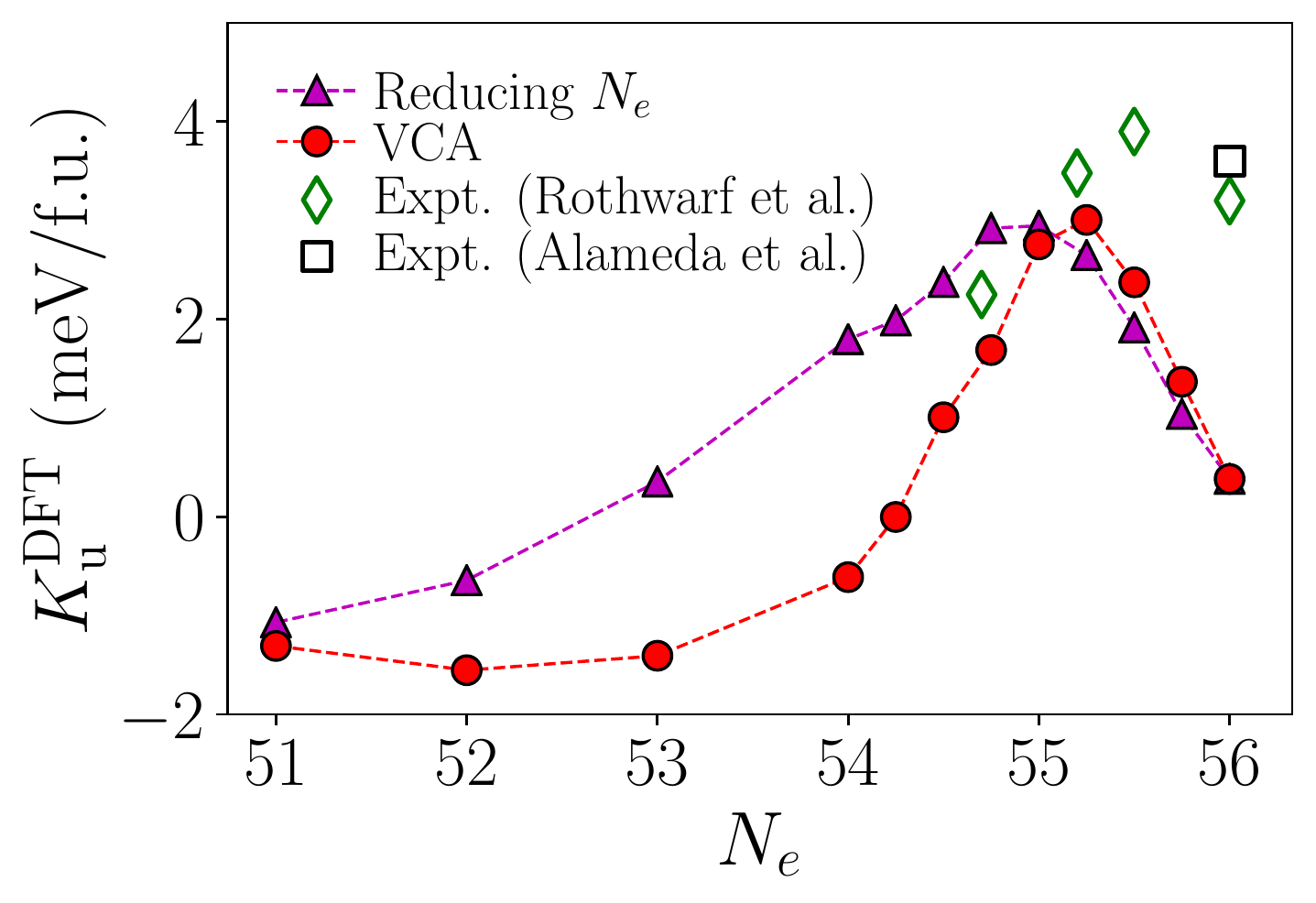}
 \caption{Calculated MCA energy, $K_{\mathrm{u}}^{\mathrm{DFT}}$, as the function of valance electron number obtained by manually reducing $N_e$ and VCA approaches, respectively. The computational results are compared with the experimental values reported by Rothwarf \textit{et al.}~\cite{YCoFe5_mae_exp} and Alameda \textit{et al.}~\cite{YCo5_mae_exp}. The dash lines are shown to guide the eye. }
 \label{fig:K_u_VCA}
\end{figure}

On the basis of the above analyses, we now discuss possible scenarios explaining the inconsistency between our computational result and experiments about the stability of the CaCu$_5$-type YCo$_5$. 
First, since the detailed structure analyses using XRD measurement have been reported only at room temperature so far, there could be a lower-symmetry phase of pristine YCo$_5$ in the low-temperature region which is left to be discovered. Should there be a structural phase transition, it would be second-order because no anomaly has been observed in the lattice parameters down to 5 K~\cite{YCo5_thermal}.
The second scenario is the possible off-stoichiometry in the experimental samples. YCo$_5$ is the mother compound from which other Co-poor and Co-rich phases, including  YCo$_3$, Y$_2$Co$_7$, Y$_5$Co$_{19}$, and Y$_2$Co$_{17}$, are derived by partial substitution~\cite{YCo5_dumbbell}. Since the formation energies of these derivatives are comparable to that of YCo$_5$~\cite{Ishikawa_PRM2021}, the off-stoichiometry of YCo$_{5+z}$ can occur rather easily in the samples prepared by a high-temperature treatment. Indeed, Pareti \textit{et al.}~\cite{YCo5+z_exp} reported that YCo$_{5+z}$ could be synthesized for the range of $-0.7<z<2.3$ by performing a quenching from high temperature. The Co-poor off-stoichiometry introduces excess holes to the system, which are expected to decrease the population of the antibonding state at the Fermi level and thereby improve the stability of the structure.
Third, the phonon instability of YCo$_{5}$ might appear due to the limited accuracy of the present HLD calculation based on GGA-PBE. While we have also observed the same phonon instability with other semilocal functionals such as PBEsol, more accurate treatment of the electronic correlation, for example, by hybrid functionals or by the combination of the dynamical mean-field theory and the DFT (DFT+DMFT) may yield stable phonons. Moreover, the quantum fluctuation of atomic nuclei may help avoid the structural distortion to occur and stabilize the CaCu$_5$-type YCo$_5$, akin to the case of a superconducting hydride~\cite{Errea_Nature2020}. Answering which of these scenarios is the most plausible would be challenging as it requires additional experimental and theoretical investigations, which are left for a future study.

\subsection{MCA energy of YCo$_{5-x}$}
\label{subsec:MAE_constant}

In the following, we will focus on the MCA energy of pristine YCo$_5$ ($N_e=56$) as well as the hole-doped systems ($N_e < 56$) at 0~K and finite temperatures. In the case of results at 0~K, we compare the $K_{\mathrm{u}}^{\mathrm{DFT}}$ with $K_{\mathrm{u}}^{\mathrm{PT}}$ for various $N_e$ values. 
In addition, we also calculated MCA energy, $K_{\mathrm{u}}^{\mathrm{VCA}}$, of the Y(Fe,Co)$_5$ alloy using virtual crystal approximation (VCA)~\cite{VCA,VCA_1}
to compare with the experimental results. Note the volume of Y(Fe,Co)$_5$ alloy is fixed for the VCA calculation, similar to the case of manually reducing $N_e$. The corresponding total number of valance electrons for the VCA calculations, shown in the first column of Table~\ref{table:MAE}, is also labeled as $N_e$. In addition, the spin moment, namely $m_{\sigma,\mathrm{Co}}$, at different Co sites and anisotropy in orbital moment, defined as $\Delta{}m_{\mathrm{o}} = m_{\parallel}^{\mathrm{o}}-m_{\perp}^{\mathrm{o}}$, are also studied. 
As for the finite-temperature effects, we calculated the HLD contribution $K_{\mathrm{u}}^{\mathrm{phon}}$ [Eq.~(\ref{eq:K_vib})] for the dynamically stable cases and the ensemble average $\langle{}K_{\mathrm{u}}\rangle_{\mathrm{MD}}$ [Eq.~(\ref{eq:K_md})] for $N_e$ = 54, 55, and 56 (pristine YCo$_5$).

\subsubsection{MCA energy at zero kelvin}
\label{subsubsection:MAE_0K}

\begin{figure*}[!tbhp]
 \centering
 \includegraphics[width=\textwidth]{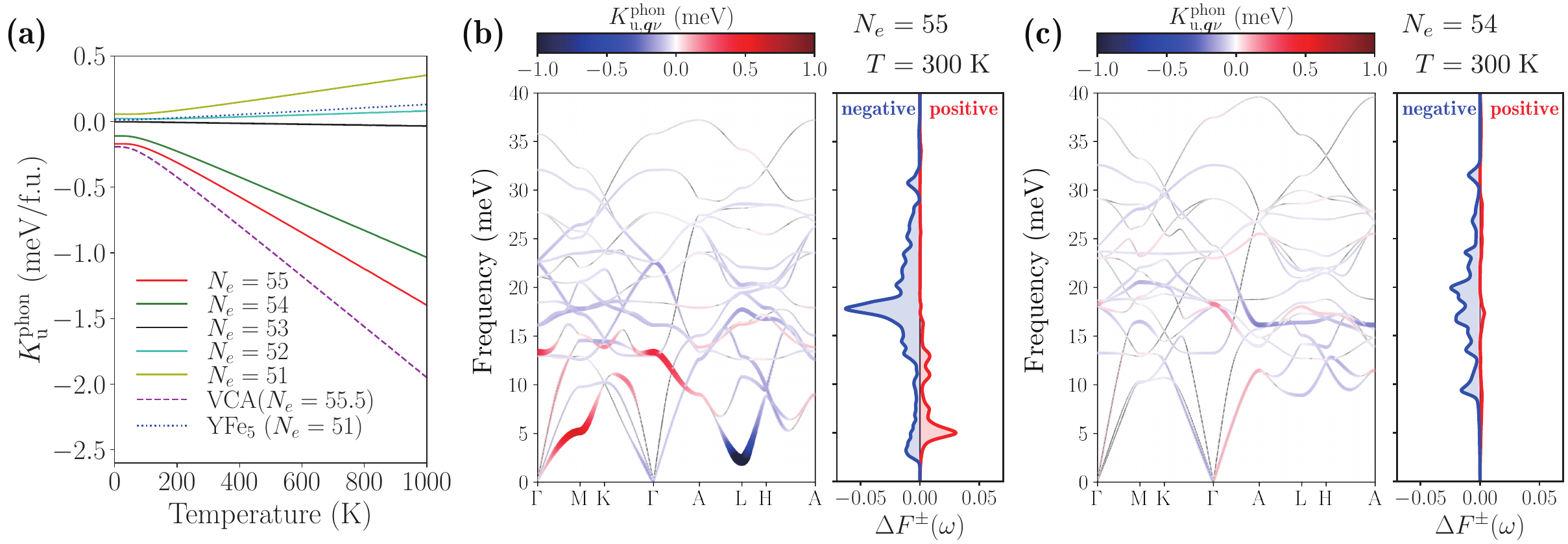}
 \caption{(a) Calculated temperature dependence of $K_{\mathrm{u}}^{\mathrm{phon}}$. Solid lines represent $K_{\mathrm{u}}^{\mathrm{phon}}$ of hole-doped YCo$_5$ obtained by manually reducing $N_e$, dashed line represents $K_{\mathrm{VCA}}^{\mathrm{phon}}$ of Y(Co,Fe)$_5$ obtained by VCA approach with $N_e$ = 55.5, and dotted line is the $K_{\mathrm{u}}^{\mathrm{phon}}$ value of YFe$_5$. (b) and (c) show the room-temperature ($300$~K) mode- and energy-decomposed $K_{\mathrm{u}}^{\mathrm{phon}}$ of hole-doped YCo$_5$ with $N_e=$ 55, and 54, respectively. In the mode-decomposed figures, the line color represents the sign of $K_{\mathrm{u},\bm{q}\nu}^{\mathrm{phon}}$, whose absolute value is represented by the linewidth.}
 \label{fig:vib-decomposed}
\end{figure*}

First, we discuss the $N_e$ dependency of the MCA energy at 0~K obtained from the different approaches mentioned above. 
The present computational results are shown in Fig.~\ref{fig:K_u_VCA} and Table~\ref{table:MAE}. We also show the $N_e$ dependence of the spin magnetic moments in Appendix~\ref{spin_mag}.
As shown in Fig.~\ref{fig:K_u_VCA} and Table~\ref{table:MAE}, all approaches gave similar trends in that positive MCA energy first increases, then decreases, and finally changes the sign to negative as $N_e$ decreases. This trend is qualitatively consistent with the MCA constants of Y(Fe$_x$Co$_{1-x}$)$_5$ ($x$ = 0.00, 0.10, 0.16, and 0.26) alloys measured by Rothwarf \textit{et al.}~\cite{YCoFe5_mae_exp}. 
Both $K_{\mathrm{u}}^{\mathrm{DFT}}$ and $K_{\mathrm{u}}^{\mathrm{PT}}$ values of the pristine YCo$_5$ ($N_e$ = 56) are smaller than the experimental values of $\sim$3.6 meV/f.u.~\cite{YCo5_mae_exp} and $\sim$3.20 meV/f.u.~\cite{YCoFe5_mae_exp1}. 
Similar underestimation problems have also been observed in the previous LDA- or GGA-level calculations~\cite{YCo5_mae_GGAU, YCo5_mag_FLAPW,YCo5_mae_DMFT}, which have successfully been resolved either by considering the orbital polarization contribution~\cite{YCo5_orbit,YCo5_mag_FLAPW,Orbital_polarization}, by using DFT+U approach~\cite{YCo5_mae_GGAU}, or by DFT+DMFT~\cite{YCo5_mae_DMFT}. Nonetheless, since the main subject of this work is to study the lattice dynamics effects on the MCA energy, and the GGA-PBE functional gives the reasonable results of the $N_e$ dependency (Fig.~\ref{fig:K_u_VCA}), we still employ the GGA-PBE functional in the study.

From the results of the second-order perturbation calculation, we found that the significant enhancement of the MCA energy at $N_e$ = 55 originates from the spin conservation term $K_{\downarrow\Rightarrow\downarrow}^{i}$ at the Co sites. Moreover, we have calculated the anisotropy of orbital moments $\Delta m_{\mathrm{o}}$ (see Table \ref{table:MAE}) and confirmed that the Bruno relation ($K_{\mathrm{u}}^{\mathrm{DFT}}\propto\Delta m_\mathrm{o}$)~\cite{Bruno_relation,Bruno_relation1} nicely holds for $N_e \geq 54$. More detailed discussion can be found in Appendix~\ref{Bruno_relation}.

\subsubsection{Lattice dynamics effects on the MCA energy at finite temperature}
\label{subsub:MAE_finite}

Next, we discuss the lattice dynamics effects on the MCA energy of the pristine and hole-doped YCo$_{5}$. Figure~\ref{fig:vib-decomposed}(a) shows the temperature dependence of $K_{\mathrm{u}}^{\mathrm{phon}}$ [Eq.~(\ref{eq:K_vib})] calculated for the hole-doped YCo$_{5}$ ($N_e \leq 55)$, for which the phonons are dynamically stable at 0 K. In the $N_e \geq 54$ region, the $K_{\mathrm{u}}^{\mathrm{phon}}$ value gradually decreases as the temperature raises. As $N_e$ decreases further, the temperature dependence becomes weaker, and the sign of the slope eventually changes at $N_e \simeq 53$. To understand the origin of this $N_e$ dependence, we computed the mode- and energy-decomposed $K_{\mathrm{u}}^{\mathrm{phon}}$ for $N_e = 55$ and 54, as shown in Fig.~\ref{fig:vib-decomposed}(b), and (c), respectively. Here, the mode-decomposed value was defined as $K_{\mathrm{u},\bm{q}\nu}^{\mathrm{phon}}=\frac{\hbar}{2} [1 + 2n(\omega_{\bm{q}\nu,\parallel})] \Delta \omega_{\bm{q}\nu}$, and the energy decomposition was performed by 
computing
\begin{equation}
    \Delta F^{\pm}(\omega) = \frac{1}{N_q}\sum_{\bm{q}\nu}K_{\mathrm{u},\bm{q}\nu}^{\mathrm{phon}} \delta(\omega-\omega_{\bm{q}\nu,\parallel}) \theta(\pm \Delta\omega_{\bm{q}\nu}),
\end{equation}
where $\theta(x)$ is the step function which becomes $\theta(x) = 1$ when $x>0$ and 0 otherwise.
$\Delta F^{+}(\omega)$ ($\Delta F^{-}(\omega)$) includes the contributions from phonon modes that increase (decrease) $K_{\mathrm{u}}^{\mathrm{phon}}$. It is straightforward to show $\int_{0}^{\infty}d\omega [\Delta F^{+}(\omega) + \Delta F^{-}(\omega)]\approx K_{\mathrm{u}}^{\mathrm{phon}}$. It is remarkable in Fig.~\ref{fig:vib-decomposed}(b) that the soft phonon at the L point contributes negatively to the MCA energy at finite temperature. On the other hand, the lower transverse acoustic mode along the $\Gamma$-M line and the second-lowest optical mode around the $\Gamma$ point have relatively large positive contributions. Moreover, $\Delta F^{-}(\omega)$ shows a large negative peak around $\omega\sim 18$ meV, which can be attributed to the weakly dispersive branches in this energy region. 
When one more electron is removed from the system, the antibonding nature further weakens, and consequently, the overall phonon frequency increases, as shown in Fig.~\ref{fig:vib-decomposed}(c). This hardening decreases $n(\omega_{\bm{q}\nu,\parallel})$ in Eq.~(\ref{eq:K_phon}), which partially explains the disappearance of the clear structures in the mode- and energy-resolved $K_{\mathrm{u}}^{\mathrm{phon}}$. 
We also observed the same qualitative change in the VCA calculation, which is shown in Fig.~S3 of the SM~\cite{supplement}. 

The zero-temperature MCA energy, shown in Fig.~\ref{fig:K_u_VCA}, takes the maximum value at around $N_e=55$. By contrast, the lattice dynamics contribution tends to become more significant as increasing $N_e$ from 54 to 56. Hence, the effect of the lattice dynamics on the MCA energy is expected to be larger in the large $N_e$ region close to $N_e=56$. Indeed, in the VCA calculation with $N_e=55.5$, the $K_{\mathrm{u}}^{\mathrm{phon}}$ value at 800 K is around $-1.5$ meV/f.u. (dashed line in Fig.~\ref{fig:vib-decomposed}(a)), whose absolute value is comparable to that of $K_{\mathrm{u}}^{\mathrm{DFT}}=2.4$ meV/f.u. Accordingly, we may expect that the effect of the lattice dynamics would become even more significant for the pristine YCo$_{5}$ ($N_e = 56$). However, as $N_e$ approaches 56, the phonon frequencies tend to soften, especially noticeable at the L point, and the effect of phonon anharmonicity would be greater. The phonon anharmonicity is expected to influence the phonon frequencies at finite temperatures and thereby the Bose--Einstein occupation function, which would affect the temperature-dependence of $K_{\mathrm{u}}^{\mathrm{phon}}$. Moreover, the (incipient) distortion of the hexagonal structure, which is not considered in the HLD approach, may also influence temperature dependence.

\begin{figure}[!tb]
 \centering
 \includegraphics[width=0.48\textwidth]{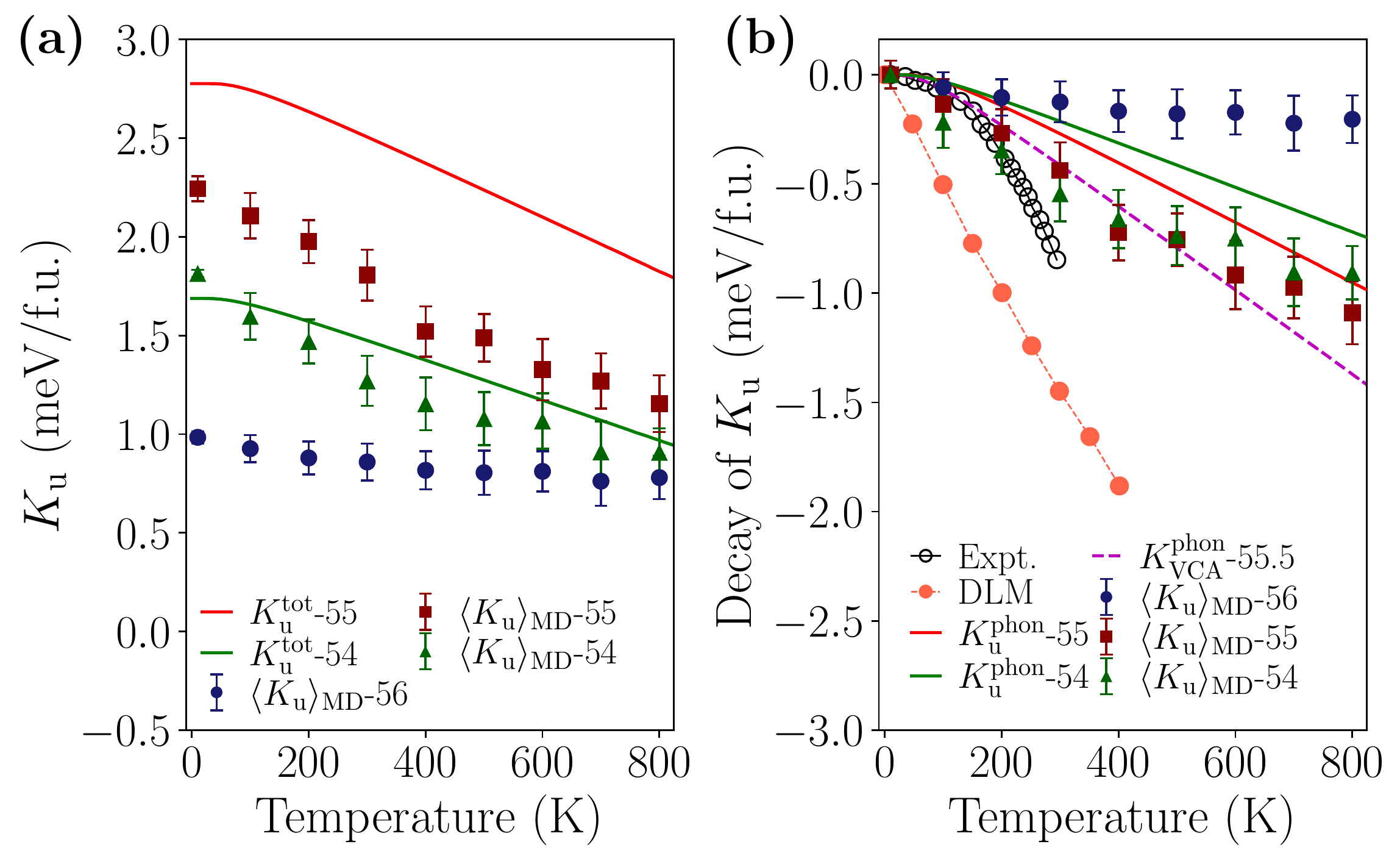}
 \caption{Temperature dependence of MCA energy evaluated with different methods. (a) Filled symbols are $\langle{}K_{\mathrm{u}}\rangle_{\mathrm{MD}}$ values of YCo$_5$ evaluated from AIMD with $N_e$ = 56, 55, and 54, respectively. Solid lines are $K_{\mathrm{u}}^{\mathrm{tot}}$ (see text) of YCo$_5$ with $N_e$ = 55, and 54, respectively. (b) Temperature decay of $K_{\mathrm{u}}$ for YCo$_5$ with $N_e$ = 56, 55, 55.5, and 54, respectively. Dash line is the $K_{\mathrm{VCA}}^{\mathrm{phon}}$ of Y(Co,Fe)$_5$ ($N_e=55.5$) determined using VCA method. The  experimental data (black open circles) and theoretical results (red solid circles) based on disordered local moment (DLM) method shown were taken from Ref.~\cite{YCo5_mae_exp} and Ref.~\cite{YCo5_mae_DLM}, respectively. The error bars are the calculated standard deviation obtained with the sampling structures.}
 \label{fig:vib-AIMD}
\end{figure}

To understand the effect of the phonon anharmonicity and the possible structural distortion on the MCA energy at finite temperature, we compare the calculated $\braket{K_{\mathrm{u}}}_{\mathrm{MD}}$ values with $K_{\mathrm{u}}^{\mathrm{tot}}=K_{\mathrm{u}}^{\mathrm{DFT}}+ K_{\mathrm{u}}^{\mathrm{phon}}$ in Fig.~\ref{fig:vib-AIMD}(a). These two values should be similar in the high-temperature (classical) limit when phonon anharmonicity is weak. Besides, when there are no structural distortion, $\braket{K_{\mathrm{u}}}_{\mathrm{MD}}\approx K_{\mathrm{u}}^{\mathrm{DFT}}$ should hold in the $T\rightarrow 0$ limit because the present AIMD neglects the nuclear quantum effect, i.e., zero-point motion. Indeed, for $N_e=54$ where the phonon frequencies are relatively higher than the $N_e=55$ case, $\braket{K_{\mathrm{u}}}_{\mathrm{MD}}$ and $K_{\mathrm{u}}^{\mathrm{tot}}$ agree reasonably well with each other. Also, the $\braket{K_{\mathrm{u}}}_{\mathrm{MD}}$ value at 10 K is 1.815 meV/f.u., which is in close agreement with $K_{\mathrm{u}}^{\mathrm{DFT}}=1.797$ meV/f.u. (see Table \ref{table:MAE}). By contrast, in the case of $N_e=55$, the  $\braket{K_{\mathrm{u}}}_{\mathrm{MD}}$ value at 10 K is $\sim$2.24 meV/f.u., which is $\sim$24\% smaller than $K_{\mathrm{u}}^{\mathrm{DFT}}=2.947$ meV/f.u. We also observed similar discrepancy between the canonical average of the orbital moment, $\braket{\Delta m_{\mathrm{o}}}_{\mathrm{MD}}$, at 10 K and the $\Delta m_{\mathrm{o}}$ value obtained for the perfect $P6/mmm$ structure shown in Table \ref{table:MAE} (see Fig.~S4 in the SM~\cite{supplement}). These discrepancies indicate that the hexagonal $P6/mmm$ structure transformed to another lower-energy structure in the present AIMD simulations at low temperatures even though the harmonic phonon was dynamically stable when $N_e=55$ (Fig.~\ref{fig:phonon-main}(a)). We have found that another structure possessing the $P6_3/mcm$ symmetry was slightly more stable than the $P6/mmm$ phase by $\sim$0.25 meV/f.u., which explains the observed discrepancy. Nevertheless, except for the offset at $T = 0 $ K, the temperature decay of $\braket{K_{\mathrm{u}}}_{\mathrm{MD}}$ is quite similar to that of $K_{\mathrm{u}}^{\mathrm{tot}}$ in the wide temperature range. Hence, the effect of the anharmonicity on the MCA energy is insignificant for $N_e=55$.

\begin{figure}[!tb]
 \centering
 \includegraphics[width=0.4\textwidth]{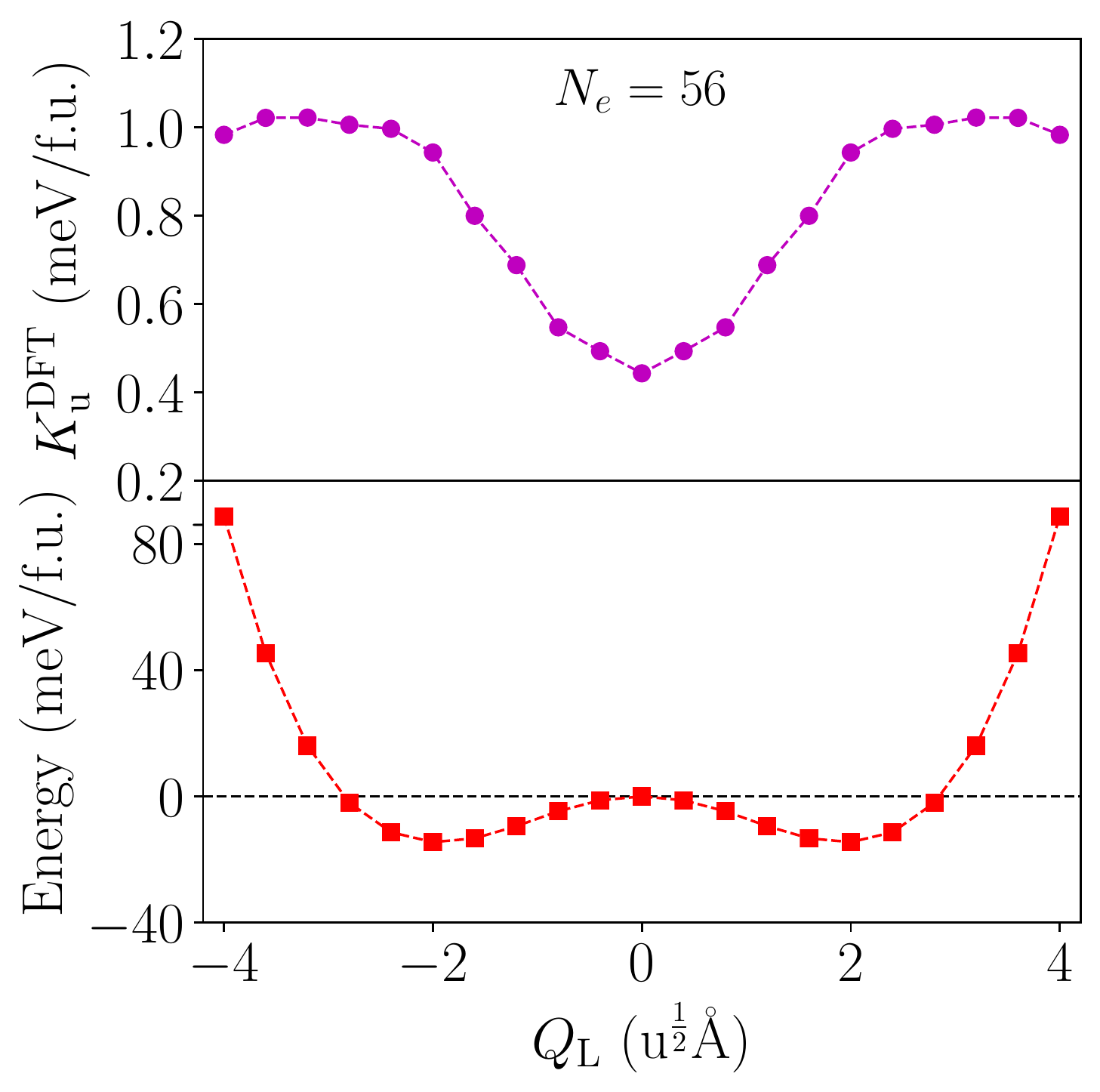}
 \caption{Calculated MCA energy, $K_{\mathrm{u}}^{\mathrm{DFT}}$, (upper panel) and total energy (lower panel) as the function of normal coordinate amplitude $Q_{\mathrm{L}}$ of the L-point soft mode for the pristine YCo$_5$. Note the total energy has been offset by the corresponding value at $Q_{\mathrm{L}}=0$. The dash lines are shown to guide the eye. }
 \label{fig:K_u_Q}
\end{figure}

It is interesting to see that, in the case of $N_e=56$, the $\braket{K_{\mathrm{u}}}_{\mathrm{MD}}$ values were larger than $K_{\mathrm{u}}^{\mathrm{DFT}}=0.383$ meV/f.u. both in low- and high-temperature ranges. Also, the $\braket{K_{\mathrm{u}}}_{\mathrm{MD}}$ value is almost temperature independent. To obtain deeper insights into the mechanism behind them, we calculated the $K_{\mathrm{u}}^{\mathrm{DFT}}$ value as a function of the normal coordinate amplitude $Q_{\mathrm{L}}$ of the L-point soft mode. As shown in Fig.~\ref{fig:K_u_Q}, the sign of $\partial^2 K_{\mathrm{u}}^{\mathrm{DFT}}/\partial Q_{\mathrm{L}}^2$ at $Q_{\mathrm{L}} = 0$ is positive when $N_e=56$. Since the L-point soft mode induces the structural distortion, it should affect the temperature dependence of $\braket{K_{\mathrm{u}}}_{\mathrm{MD}}$ most significantly. Hence, we can approximately write 
\begin{equation}
    \braket{K_{\mathrm{u}}}_{\mathrm{MD}} \approx \int_{-\infty}^{\infty}K_{\mathrm{u}}^{\mathrm{DFT}}(Q_{\mathrm{L}})P_{T}(Q_{\mathrm{L}}) \mathrm{d}Q_{\mathrm{L}},
    \label{eq:K_MD_QL}
\end{equation}
where $P_T(Q_{\mathrm{L}})$ is the probability distribution of $Q_{\mathrm{L}}$ at temperature $T$. In the low-temperature region, $P_T(Q_{\mathrm{L}})$ has a peak around either side of the double-well minima, where $K_{\mathrm{u}}^{\mathrm{DFT}}(Q_{\mathrm{L}})$ becomes $\sim$1.0 meV/f.u. This value is in good agreement with $\braket{K_{\mathrm{u}}}_{\mathrm{MD}}$ at 10 K. In the high-temperature limit, $P_T(Q_{\mathrm{L}})$ has a peak around $Q_{\mathrm{L}}=0$, i.e., $\braket{Q_{\mathrm{L}}}\approx 0$, but the variance of the distribution $\braket{Q_{\mathrm{L}}^2}$ becomes large. Thus,  the $\braket{K_{\mathrm{u}}}_{\mathrm{MD}}$ value became larger than $K_{\mathrm{u}}^{\mathrm{DFT}}(0)$ because of $\partial^2 K_{\mathrm{u}}^{\mathrm{DFT}}/\partial Q_{\mathrm{L}}^2 >0$. We note that, for $N_e=55$ and 54, the sign of $\partial^2 K_{\mathrm{u}}^{\mathrm{DFT}}/\partial Q_{\mathrm{L}}^2$ is negative (see Fig.~S5 in the SM~\cite{supplement}), which is consistent with the negative sign of $K_{\mathrm{u},\bm{q}\nu}^{\mathrm{phon}}$ shown in Figs.~\ref{fig:vib-decomposed}(b), (c).

We also show $\braket{K_{\mathrm{u}}}_{\mathrm{MD+TE}} = \braket{K_{\mathrm{u}}}_{\mathrm{MD}} + K_{\mathrm{u}}^{\mathrm{TE}}$ (see Fig.~S6 in the SM~\cite{supplement}) which includes the effect of thermal expansion evaluated approximately as explained in Sec.~\ref{subsec:MAE_finite}. It is clear that the effect of thermal expansion is insignificant except for the $N_e=54$ case at high temperatures, where the $K_{\mathrm{u}}^{\mathrm{TE}}$ value amounts to $-0.3$ mev/f.u.

Figure \ref{fig:vib-AIMD}(b) compares the predicted temperature decay of the MCA energy obtained from various computational methods, including the DLM result of Ref.~\cite{YCo5_mae_DLM}, and that of the experimental result. Note that the data shown in Fig.~\ref{fig:vib-AIMD}(b) have been offset by the corresponding value at the lowest temperature: 10~K for $\langle{}K_{\mathrm{u}}\rangle_{\mathrm{MD}}$, and 0 K for the others. The present calculation based on GGA-PBE predicts that lattice dynamics hardly influences $K_{\mathrm{u}}$ for the pristine YCo$_5$ ($N_e=56$) at finite temperatures mainly due to the presence of the structural distortion. Therefore, it is reasonable to conclude that the lattice dynamics effect on the MCA energy is negligible for the pristine YCo$_5$ and the spin fluctuation remains the dominant effect for yielding the temperature decay of $K_{\mathrm{u}}$. However, when the structural distortion is suppressed by hole-doping, the lattice dynamics effect becomes non-negligible in the $N_e\gtrsim54$ region and therefore should be considered as well when comparing theoretical results with experimental ones.

\section{Summary}

\label{sec:summary}

To summarize, we theoretically studied the structural stability and lattice dynamics effects on the MCA energy for pristine and hole-doped YCo$_5$ using first-principles methods based on DFT. The CaCu$_5$-type ($P6/mmm$) structure of pristine YCo$_5$ is predicted to be dynamically unstable at 0 K, and the soft phonon modes appear at L and M points. This phonon instability originates from the large population of antibonding states near the Fermi level which is formed by the nearest Co atoms in the honeycomb layer. We demonstrated that the phonon instability can be removed by hole doping, which depopulates the antibonding states. 
We also evaluated the effect of lattice dynamics on the MCA energy based on harmonic lattice dynamics and AIMD methods by computing $K_{\mathrm{u}}^{\mathrm{phon}}$ and $\langle{}K_{\mathrm{u}}\rangle_{\mathrm{MD}}$, respectively. For the pristine YCo$_5$, we found that lattice dynamics hardly influence the MCA energy at finite temperature. Also, the ensemble average $\braket{K_{\mathrm{u}}}_{\mathrm{MD}}$ at finite temperature turned out to be larger than the $K_{\mathrm{u}}^{\mathrm{DFT}}$ computed with the static $P6/mmm$ lattice. We attributed these unique features to the dynamical distortion of the structure, which is induced by the soft phonon modes. 
By contrast, in the hole-doped YCo$_5$ where the dynamical distortion is suppressed, a much larger temperature decay was observed both in $K_{\mathrm{u}}^{\mathrm{phon}}$ and $\braket{K_{\mathrm{u}}}_{\mathrm{MD}}$. While the lattice dynamic effect on the finite-temperature MCA energy is not as significant as that of the spin fluctuation, its effect should not be neglected when making a quantitative comparison between theory and experiments, particularly for the hole-doped YCo$_5$.

\begin{acknowledgments}
We thank K. Masuda for the fruitful discussion and X. He for assistance with the COHP analysis. This work was partially supported by the Ministry of Education, Culture, Sports, Science and Technology (MEXT) as ``Elements Strategy Initiative Center for Magnetic Materials'' (ESICMM, Project ID: JPMXP0112101004) and as ``Program for Promoting Researches on the Supercomputer Fugaku'' (DPMSD, Project ID: JPMXP1020200307). The figures of crystal structures are created by the \textsc{vesta} software~\cite{vesta}.
\end{acknowledgments}

\renewcommand{\appendixname}{APPENDIX}

\appendix
\section{MAGNETIC MOMENT }
\label{spin_mag}

\begin{table}[!bt]
\begin{threeparttable}
\centering
  \caption{Calculated spin magnetic moment for different Co sites ($m_{\sigma,\mathrm{Co}}$). The spin magnetic moments of previous works (experimental or theoretical) are obtained from Refs.~\cite{YCo5_mag_exp,YCo5_mag_exp_1,YCo5_latt_exp,YCo5_mae_GGAU,YCo5_mag_FLAPW,YFe5_mag_GGA}}
  \label{table:spin_moment}
  \small
  \begin{ruledtabular}
  \begin{tabular}{lcccc}
    \multirow{3}{*}{$N_e$} &\multicolumn{4}{c}{$m_{\sigma,\mathrm{Co}}$ ($\mu_{\mathrm{B}}$)} \\
    \cline{2-5}
     &\multicolumn{2}{c}{This study}   &\multicolumn{2}{c}{Previous work} \\
     \cline{2-5}
     &   Co$_{2c}$ & Co$_{3g}$  &Co$_{2c}$   &Co$_{3g}$ \\
    \toprule[0.15mm]
     \multirow{2}{*}{56} &\multirow{2}{*}{1.498}   &\multirow{2}{*}{1.525}  &1.44\tnote{a}\hspace{2mm}1.62\tnote{b}   &1.31\tnote{a}\hspace{2mm}1.64\tnote{b}\\
        &  &      &1.61\tnote{c}\hspace{2mm}1.46\tnote{d}&1.68\tnote{c}\hspace{2mm}1.51\tnote{d} \\
     55 &1.685   &1.668&&\\
     54 &1.833   &1.810&&\\
     53 &1.950   &1.953&&\\
     52 &2.061   &2.100&&\\
     51 &2.170   &2.231&2.05\tnote{e}&2.00\tnote{e}\\
  \end{tabular}
  \end{ruledtabular}
  \begin{tablenotes}[para,flushleft]
  \item[a] Experimental results, Ref.~\cite{YCo5_mag_exp,YCo5_mag_exp_1}
  \item[b] DLM, Ref.~\cite{YCo5_latt_exp}
  \item[c] GGA+U,Ref.~\cite{YCo5_mae_GGAU}
  \item[d] FLAPW, Ref.~\cite{YCo5_mag_FLAPW}
  \item[e] YFe$_5$,GGA, Ref.~\cite{YFe5_mag_GGA} 
  \end{tablenotes}
\end{threeparttable}
\end{table}

\begin{figure}[t]
 \centering
 \includegraphics[width=0.46\textwidth]{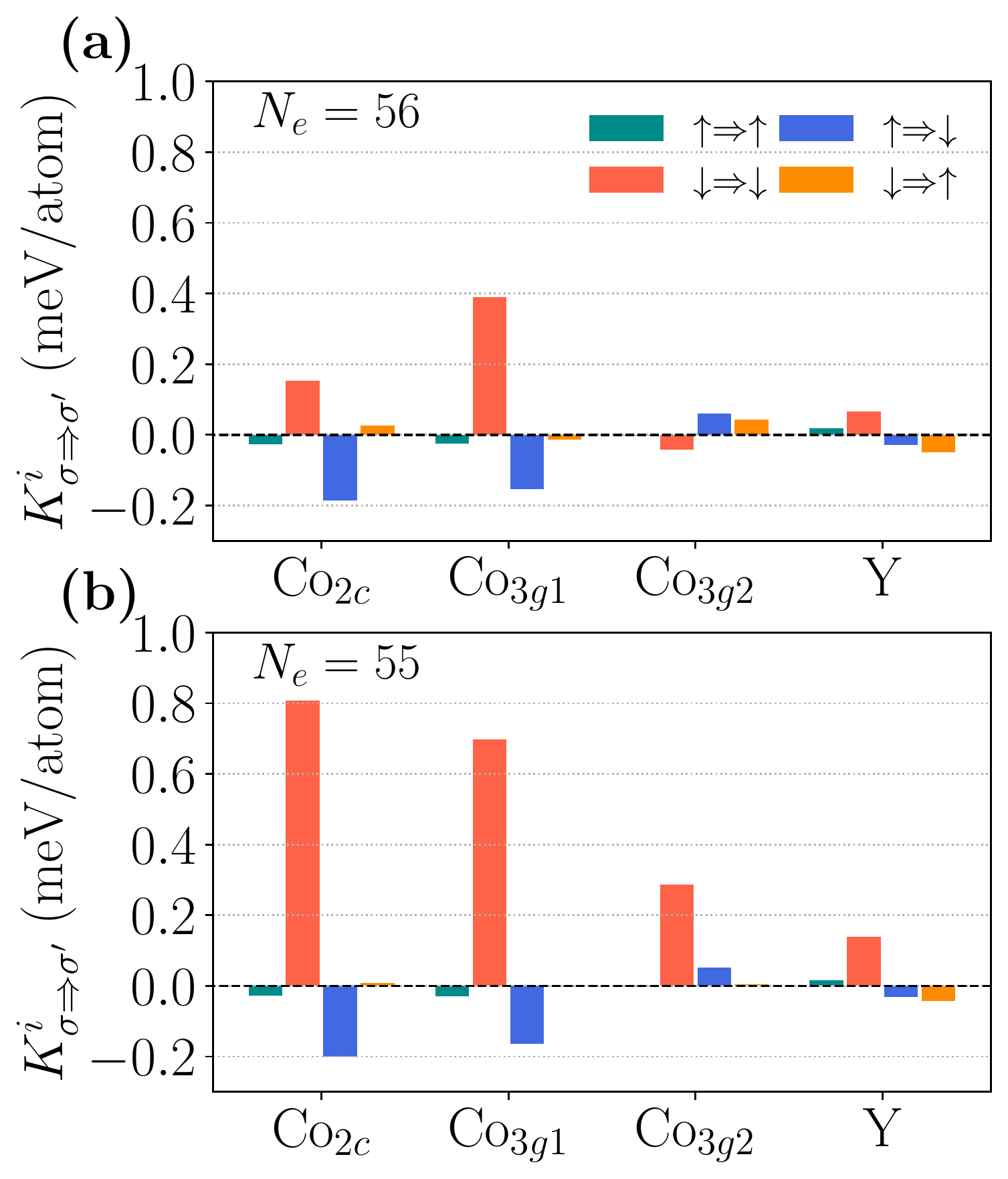}
 \caption{Decomposed atomic sites dependence of MCA energy, $K_{\sigma\Rightarrow\sigma'}^{i}$, obtained from second-order perturbation analysis in YCo$_5$ with $N_e$ = 56 (a) and 55 (b), respectively.}
 \label{fig:second_order}
\end{figure}

 The $N_e$-dependent spin magnetic moments at different Co sites, $m_{\sigma,\mathrm{Co}}$, are listed in Table~\ref{table:spin_moment}. For $N_e$ = 56, the calculated spin moments of Co$_{2c}$ and Co$_{3g}$ are 1.498 and 1.525 $\mu_\mathrm{B}$, respectively, which is slightly larger than the experimental values of 1.44 and 1.31 $\mu_{\mathrm{B}}$~\cite{YCo5_mag_exp,YCo5_mag_exp_1}. To compare with previous theoretical results, a selection of results based on different methods is summarized in Table~\ref{table:spin_moment}. These values are spin moments obtained by DLM method~\cite{YCo5_latt_exp}, GGA+U method~\cite{YCo5_mae_GGAU}, full potential linearized augmented plane wave (FLAPW) method~\cite{YCo5_mag_FLAPW}. Our results match well with the FLAPW values but differ slightly from the GGA+U and DLM results. The spin moments of both Co sites increase with decreasing $N_e$, which can be explained by the DOS shown in Fig.~S7(a), and (b) of the SM~\cite{supplement}. When $N_e$ is decreased, the minority spin states shift to the higher energy, increasing spin splitting. Since the majority spin states are fully occupied, the decrease of minority spin states will lead to an increase of the net single spin states, then a larger spin magnetic moment. Note that the spin magnetic moments for $N_e$ = 51 is slightly larger than that of YFe$_5$ with the values of 2.082, and 2.033 $\mu_\mathrm{B}$ for Fe$_{2c}$ and Fe$_{3g}$ according to our calculations, which may be attributed to the not fully occupied $3d$ majority spin states (see Fig.~S7(d) in the SM~\cite{supplement}) in YFe$_5$ ($N_e=51$). As shown in Fig.~S7(c) and (d) of the SM~\cite{supplement}, the majority spin states of Y(Fe,Co)$_5$ evaluated by the VCA approach will become partially occupied when $N_e\leq52$, which is different from the DOS determined by reducing $N_e$ approach. 

\section{VALIDITY OF THE BRUNO RELATION}
\label{Bruno_relation}

According to the Bruno relation, the MCA energy is proportional to the anisotropy in the orbital moment ($K_{\mathrm{u}}^{\mathrm{DFT}}\propto\Delta m_\mathrm{o}$). To understand the validity of the Bruno relation with $N_e$, the atomic site dependence of $K_{\sigma\Rightarrow\sigma'}^{i}$ with different spin-transition processes are shown in Fig.~\ref{fig:second_order} and Fig.~S8 of the SM~\cite{supplement}, respectively. 
The maximum $K_{\mathrm{u}}^{\mathrm{PT}}$ for $N_e$ = 55 originates mainly from the enhancement of positive spin conservation term $K_{\downarrow\Rightarrow\downarrow}^{i}$ both at Co$_{2c}$, Co$_{3g1}$, and Co$_{3g2}$ sites. However, positive $K_{\downarrow\Rightarrow\downarrow}^{i}$ decrease dramatically as NE decreases (see Fig.~S8 in the SM~\cite{supplement}). The negative spin-flip term $K_{\uparrow\Rightarrow\downarrow}^{i}$ of Co$_{2c}$ and Co$_{3g1}$ mainly contributes to $K_{\mathrm{u}}^{\mathrm{PT}}$ when $N_e\leq52$, resulting in a negative $K_{\mathrm{u}}^{\mathrm{PT}}$. Since the spin-flip term is absent in the orbital magnetic moment, the Bruno relation holds for $N_e\geq53$ and not for $N_e\leq52$. Although we find a qualitatively similar $K_{\mathrm{u}}^{\mathrm{PT}}$ for YCo$_5$ with $N_e$ = 51 and YFe$_5$ with the corresponding values of $-0.833$ meV/f.u., and $-0.919$ meV/f.u., respectively, the mechanism is different for these two systems. The negative $K_{\uparrow\Rightarrow\downarrow}^{i}$ (see Fig.~S8(d) in the SM~\cite{supplement}) at Co$_{2c}$ and Co$_{3g1}$ makes a main contribution to $K_{\mathrm{u}}^{\mathrm{PT}}$ for YCo$_5$ with $N_e$ = 51, while negative $K_{\downarrow\Rightarrow\downarrow}^{i}$ (see Fig.~S9 in the SM~\cite{supplement}) at Fe$_{2c}$ and Fe$_{3g1}$ sites also makes a considerable contribution besides $K_{\uparrow\Rightarrow\downarrow}^{i}$ in YFe$_5$.

\bibliography{YCo5_better.bib}

\end{document}